\documentclass[twocolumn]{hefat2020}

\ifx\pdfpagewidth\undefined
\else
\fi

\usepackage{amsmath}
\usepackage{graphicx}

\usepackage[backend=biber, 
style=ieee, isbn=false, url=false, doi=false, date=year, abbreviate=true, minbibnames=1, maxbibnames=1
]{biblatex}
 
\newcommand{\Vcphi}{\ensuremath{3.46}}

\newcommand{\VcphiAl}{\ensuremath{3.10}}

\newcommand{\VcphiTi}{\ensuremath{2.24}}

\newcommand{\VcphiCuO}{\ensuremath{4.57}}
\newcommand{\cphiCuO}{\ensuremath{C_\varphi=\VcphiCuO}}
\newcommand{\VcphiCNT}{\ensuremath{2.85}}

\newcommand{\Vcs}{\ensuremath{0.05}}

\newcommand{\VcsAl}{\ensuremath{0.45}}
\newcommand{\csAl}{\ensuremath{C_\text{S}=\VcsAl}}
\newcommand{\VcsTi}{\ensuremath{0.37}}
\newcommand{\csTi}{\ensuremath{C_\text{S}=\VcsTi}}
\newcommand{\VcsCuO}{\ensuremath{0.58}}

\newcommand{\VcsCNT}{\ensuremath{0.01}}

\newcommand{\Vct}{\ensuremath{0.23}}

\newcommand{\VctAl}{\ensuremath{0.09}}

\newcommand{\VctTi}{\ensuremath{0.01}}

\newcommand{\VctCuO}{\ensuremath{0.16}}

\newcommand{\VctCNT}{\ensuremath{0.70}}
\newcommand{\ctCNT}{\ensuremath{C_\text{T}=\VctCNT}}

\newcommand{\Vcgen}{\ensuremath{1.03}}

\newcommand{\VcAl}{\ensuremath{0.99}}

\newcommand{\VcTi}{\ensuremath{1.00}}

\newcommand{\VcCuO}{\ensuremath{0.99}}

\newcommand{\VcCNT}{\ensuremath{1.06}}

\newcommand{\Vbphi}{\ensuremath{0.55}}
\newcommand{\bphi}{\ensuremath{\beta_\varphi=\Vbphi}}
\newcommand{\VbphiAl}{\ensuremath{0.70}}

\newcommand{\VbphiTi}{\ensuremath{0.71}}

\newcommand{\VbphiCuO}{\ensuremath{0.94}}

\newcommand{\VbphiCNT}{\ensuremath{0.71}}

\newcommand{\Vbs}{\ensuremath{0.03}}
\newcommand{\bs}{\ensuremath{\beta_\text{S}=\Vbs}}
\newcommand{\VbsAl}{\ensuremath{0.36}}

\newcommand{\VbsTi}{\ensuremath{0.35}}

\newcommand{\VbsCuO}{\ensuremath{0.03}}

\newcommand{\VbsCNT}{\ensuremath{0.03}}

\newcommand{\Vbt}{\ensuremath{0.08}}
\newcommand{\bt}{\ensuremath{\beta_\text{T}=\Vbt}}
\newcommand{\VbtAl}{\ensuremath{0.04}}

\newcommand{\VbtTi}{\ensuremath{0.01}}

\newcommand{\VbtCuO}{\ensuremath{0.15}}

\newcommand{\VbtCNT}{\ensuremath{0.28}}

\newcommand{\Vrgen}{\ensuremath{0.30}}

\newcommand{\VrAl}{\ensuremath{0.54}}

\newcommand{\VrTi}{\ensuremath{0.69}}

\newcommand{\VrCuO}{\ensuremath{0.80}}

\newcommand{\VrCNT}{\ensuremath{0.50}}

\newcommand{\Vn}{\ensuremath{1167}}
\newcommand{\n}{\ensuremath{{N=\Vn}}}
\newcommand{\VnAl}{\ensuremath{292}}
\newcommand{\nAl}{\ensuremath{N=\VnAl}}
\newcommand{\VnTi}{\ensuremath{105}}
\newcommand{\nTi}{\ensuremath{N=\VnTi}}
\newcommand{\VnCuO}{\ensuremath{61}}
\newcommand{\nCuO}{\ensuremath{N=\VnCuO}}
\newcommand{\VnC}{\ensuremath{76}}
\newcommand{\nC}{\ensuremath{N=\VnC}}

\title{ 
STATISTICAL ANALYSIS OF THERMAL CONDUCTIVITY\\
EXPERIMENTALLY MEASURED IN ETHYLENE GLYCOL-BASED NANOFLUIDS\\
}

\author{Tielke J.*  and Avila, M.\\
*Author for correspondence\\
Center of Applied Space Technology and Microgravity\\
University of Bremen,\\
Bremen, 28359,\\
Germany,\\
E-mail: julia.tielke@zarm.uni-bremen.de
}

\begin{document}

\maketitle

\section*{ABSTRACT}

 We collected literature data of thermal conductivity experimentally measured in ethylene glycol-based nanofluids and investigated the influence of  concentration, temperature and nanoparticle size. We implemented statistical linear regression analysis of all data points and examined four separate nanoparticle materials – alumina, titania, copper oxide and carbon-nanotubes. We found that the statistical correlations are in good agreement with Maxwell’s effective medium theory, despite large scatter in the data. The thermal conductivity increases linearly with concentration, and in the case of carbon-nanotubes with temperature, whereas the nanoparticle size shows significant influence for alumina and titania. The large scatter in the experimental data is one of the main problems.  We suggest that there is a need for careful, detailed characterizations and measurements to quantify the potential of nanofluids. \\


\begin{table}[ht]

\renewcommand{\arraystretch}{0.8}

{\bf NOMENCLATURE}\\

{\footnotesize
\begin{tabular}{l l l}
$k_{eff}$ & [W/mK] & Effective thermal conductivity of suspension\\
$k_{f}$ & [W/mK] & Thermal conductivity base fluid\\
$k_{p}$ & [W/mK] & Thermal conductivity nanoparticle material\\
$T$	& [K] & Temperature\\
$S$ &[1/nm]	& Nanoparticle surface\\
$C$ & [-] & Regression coefficient\\
\end{tabular}

Special characters

\begin{tabular}{l l l}
$\varphi$	& [-] & Particle volume fraction ratio\\
$\beta$	& [-] & Standardized regression coefficient\\
\end{tabular}
}
\end{table}

\section*{INTRODUCTION}

 The term Nanofluids was introduced by Lee \textit{et al}. \cite{Lee1999}, who dispersed metallic nanoparticles in heat transfer fluids and measured the strongly increased thermal conductivity. A first description of the increase in thermal conductivity of a fluid by the dispersion of particles is given by Maxwell's \cite{Maxwell1881} effective medium theory
 \begin{equation}\label{eq:Maxwell}
\frac{k_\text{eff}}{k_\text{f}} = 1 + \frac{3\varphi (k_\text{p} - k_\text{f} ) }{3 k_\text{f} + ( 1 - \varphi) (k_\text{p} - k_\text{f})}.
\end{equation}
Maxwell's effective medium theory assumes homogeneously dispersed particles and gives the lower limit in the bounds of thermal conductivity derived by Hashin and Shtrikman \cite{Hashin1962}. Buongiorno \textit{et al}. \cite{Buongiorno2009} conducted a benchmark study, which showed large scatter in the experimental data due to differences in the characterization and measurements. First attempts were made to increase the quality of measurements by critically examining the different measuring techniques \cite{Keblinski2008, Eapen2010, NietodeCastro2020, Bobbo2021}, with the conclusion that insufficient characterizations were mainly responsible for scatter. However, general correlations to calculate the thermal conductivity of nanofluids were still developed resulting in various versions \cite{Eastman2004, Buongiorno2005, Mugica2020}. A first attempt using statistical linear regressions was used by Khanafer \textit{et al}. \cite{Khanafer2011}, again with the aim of having a general equation describing the increase in thermal conductivity.

In this study, we use linear regressions to quantify the influence of the parameters concentration, temperature and nanoparticle size on the experimental measured thermal conductivity. We further compare the results to Maxwell's effective medium theory for separate materials to estimate their potential. 

\section*{MATERIAL AND METHODS}

 We collected literature data and set up a database with \n \ data points of experimentally measured thermal conductivity in ethylene glycol nanofluids. We only considered studies in which the concentration, temperature and nanoparticle size are specified. The preparation of the nanofluids and method of thermal conductivity measurement are explained in each study. The database contains data for 18 different materials out of $59$ publications. First we analysed all data points together. Additionally, four separate materials are analysed, in which at least $4$ different publications with $50$ data points are necessary to ensure statistical significance. The separate analyzed materials are:
 \begin{enumeroman} 
    \item Alumina (\(Al_2O_3\), \nAl): \cite{Barbes2013, Beck2009, Eastman2001, HemmatEsfe2015, HemmatEsfe2016, Gowda2010, Kumar2018, Lee1999, Longo2013, Minakov2015, Mohammadiun2016, Murshed2012, Oh2008, Pastoriza-Gallego2011, Patel2010, Timofeeva2007, Xie2002a}
    \item Titania (\(TiO_2\), \nTi) \cite{Cabaleiro2015, Khedkar2016, Longo2013, Murshed2018, Murshed2012, Sonawane2015, Karthikeyan2008}
    \item Carbon-Nanotubes (CNT, \nC) \cite{Harish2012, Li2016, Liu2011, Shamaeil2016, Xie2003}
    \item Copper oxide (CuO, \nCuO) \cite{Barbes2014, Eastman2001, Gowda2010, Lee1999, Liu2011, Patel2010, Penas2008, Leena2015}
\end{enumeroman}

The remaining materials do not have $50$ data points out of $4$ publications and are thus not analyzed separately \cite{Agarwal2019, Akilu2017, Ali2010,Amiri2016, Eastman2001, HemmatEsfe2014, Fang2015, Guo2018, Kang2006, Kazemi2014, Kumar2016,, Madhesh2014, Mariano2013, Mariano2015, Michael2019, Moosavi2010, Murshed2012, Nikkam2017, Pastoriza-Gallego2014, Patel2010, Penas2008, Rubasingh2019, Selvam2016, Seyhan2017, Xie2002, Xie2010, Yashawantha2018, Yu2011, Yu2011a, Zyla2016, Zyla2016a, Zyla2017, Zyla2017a}.

\subsection*{Linear Statistical Model}

 We employed a linear statistical model to analyze the effect of the concentration $\varphi$, the temperature $T$ and the particle size through the specific surface $S$ on the normalized thermal conductivity 
\begin{equation}
k^*(\varphi, T, S) = \frac{k_\text{{eff}}(\varphi, T, S)}{k_\text{f}(T)}
\label{eq_1}
\end{equation}
where $k_\text{f}(T)$ is the thermal conductivity of the base fluid (ethylene glycol) as a function of the measurement of temperature. With the use of linear regressions, the predictors were implemented in the following equation with the coefficients ${C_i~(i={0, \varphi, \text{T}, \text{S}})}$ given from each regression: 
\begin{equation}\label{eq:model}
    k^*(\varphi,T,S)= C_0 + C_\varphi \,\varphi + C_\text{T}\, T^* + C_\text{S}\, S^*, 
\end{equation}
The determination of the linear model and the statistical method is fully described in a previous paper \cite{Tielke2021}, where it is applied to water-based nanofluids. 

\section*{RESULTS}

\begin{figure}[!h]
\centerline{\includegraphics[width=2.5in]{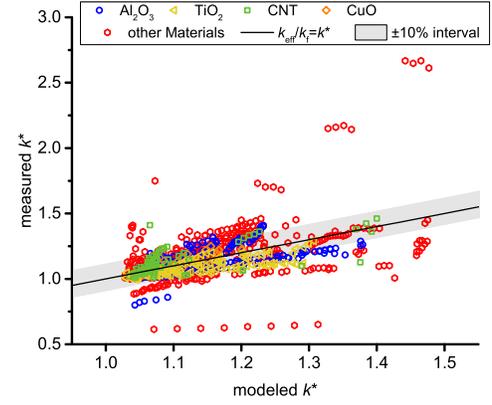}}
\caption{Experimentally measured versus modeled normalized thermal conductivity (\(k^*\)) for ethylene glycol-based nanofluids.  The ideal values of \(k^*\) and a \(\pm10\%\) interval are displayed as a solid line and a grey area, respectively. Colors are used to distinguish single materials (analyzed separately) from other materials (see the legend).} \label{Figure 2}
\end{figure}

The results of the statistical analysis of thermal conductivity modeled with the linear regression eq.~(\ref{eq:model}) for all data points and the separate materials are shown in table~\ref{tab:1} and fig.~\ref{Figure 2}, respectively. The scatter of the data, expressed by the corrected correlation coefficients $R^2 \ \text{in} \ [0.30;0.80]$ suggest that there are large uncertainties in the measurements. However, the regression for all data points is in good approximation with Maxwell's effective medium theory. The influence of each parameter, given by the standardized regression coefficients $\beta_j~(j={\varphi, \text{T}, \text{S}})$, show only a concentration-dependence with \bphi \ and insignificant coefficients towards the temperature or surface with \bt \ and \bs, respectively.  Most of the data, especially of the separately analyzed materials, lie within the $\pm 10\%$ interval, as displayed in fig.~\ref{Figure 2}. But there are materials and measurements that significantly differ from the results with modeled thermal conductivities above $k^*>1.5$.

\begin{table} [!h]
\caption {Results of the linear regressions fitted to the entire database and for separate materials. Given are the number of data points ($N$), corrected correlation coefficient ($R^2$),  model coefficients ($C_i$), standardized regression coefficients ($\beta_j$).}
\label{tab:1}
{\scriptsize
\begin{tabular} {|c|c|c|c|c c|c c|c c|}
\hline 
Material & $N$ & $R^2$ & \(C_0\) & \(C_\varphi\) & \(\beta_\varphi\) & \(C_T\) & \(\beta_T\) & \(C_S\) & \(\beta_S\) \\ \hline
all & \Vn & \Vrgen & \Vcgen & \Vcphi & \Vbphi & \Vct & \Vbt & \Vcs & \Vbs \\ \hline
$Al_20_3$ & \VnAl & \VrAl & \VcAl & \VcphiAl & \VbphiAl & \VctAl & \VbtAl & \VcsAl & \VbsAl \\ 
$TiO_2$ & \VnTi & \VrTi & \VcTi & \VcphiTi & \VbphiTi & \VctTi & \VbtTi & \VcsTi & \VbsTi \\
$CNT$ & \VnC & \VrCNT & \VcCNT & \VcphiCNT & \VbphiCNT & \VctCNT & \VbtCNT & \VcsCNT & \VbsCNT \\
$CuO$ & \VnCuO & \VrCuO & \VcCuO & \VcphiCuO & \VbphiCuO & \VctCuO & \VbtCuO & \VcsCuO & \VbsCuO \\
 \hline

\end{tabular}}
\end{table}

\begin{figure}[!h]
\centerline{\includegraphics[width=2.5in]{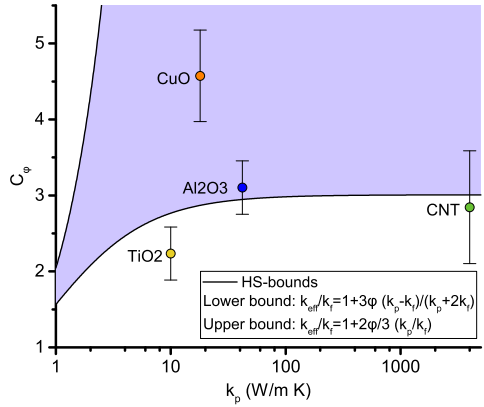}}
\caption{The colored symbols show the values of the model coefficient $C_\varphi$ (with corresponding 95\% confidence-interval, shown as error-bars) as a function of the thermal conductivity \(k_\text{p}\) of each material.The linearized HS-bounds \cite{Maxwell1881} are displayed as solid black lines, enclosing the shaded area.} \label{Figure 4}
\end{figure}

The concentration is the main parameter in the statistical correlation. In all cases, the standardized regression coefficient $\beta_\varphi$ is the highest. The modeled coefficients $C_\varphi$ are in general agreement with the the bounds derived from the linearized Maxwell equation ($C_{\varphi,~Maxwell}\in [-1.5, 3]$, see \cite{Tielke2021}). We plotted the values of $C_\varphi$ against the thermal conductivity of the particles $k_p$ in fig.~\ref{Figure 4}. We also show the bounds according to Hashin-Shtrikman \cite{Hashin1962} in their linearized form as solid black lines. The lower HS-bound is Maxwell's effective medium theory with the maximum at $C_\varphi=3$. The upper HS-bound includes all possible dispersion states. The modeled values for alumina and carbon-nanotubes are in excellent agreement with the linearized Maxwell equation, while the value for titania lies below it. If the data for higher concentrations $(\varphi \geq 2.5 vol\%)$ are excluded, the value for titania also fits to Maxwell's prediction. The value for copper oxide exceeds Maxwell's prediction, but is inside the bounds derived by Hashin-Shtrikman.

\begin{figure}[h]
\centerline{\includegraphics[width=2.5in]{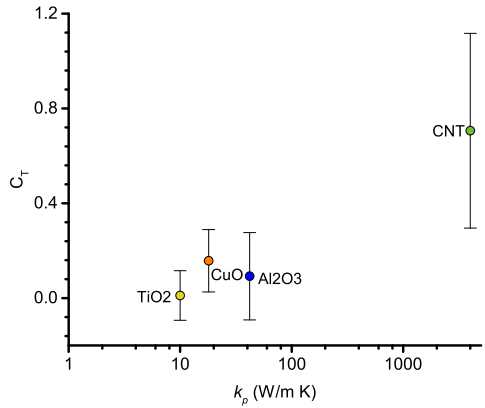}}
\caption{The colored symbols show the values of the model coefficient $C_\text{T}$ (with corresponding 95\% confidence-interval, shown as error-bars) as a function of the thermal conductivity \(k_\text{p}\) of each material.} \label{Figure 5}
\end{figure}

\begin{figure}[h]
\centerline{\includegraphics[width=2.5in]{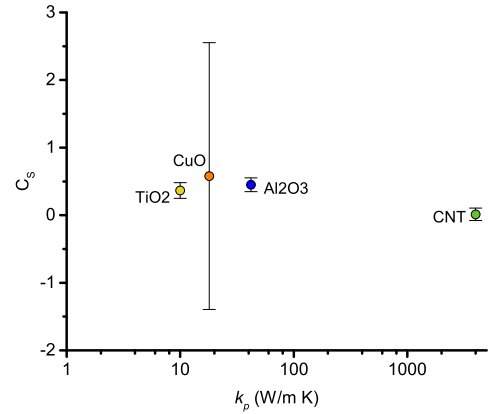}}
\caption{The colored symbols show the values of the model coefficient $C_\text{S}$ (with corresponding 95\% confidence-interval, shown as error-bars) as a function of the thermal conductivity \(k_\text{p}\) of each material.} \label{Figure 6}
\end{figure}

Similar to the concentration, we plotted the temperature- and surface-dependent coefficients $C_\text{T}$ and $C_\text{S}$ in figures~\ref{Figure 5} and \ref{Figure 6}, respectively. If the standardized regression coefficient $\beta < 0.1$, the corresponding parameter rendered insignificant. In the case of the temperature coefficient, results gave insignificant coefficients for alumina with $\beta_\text{T}=0.04$ and titania with $\beta_\text{T}=0.01$. We found a marginally significant temperature-dependence for copper oxide with $C_\text{T}=0.16 \ (\beta_\text{T}=0.15)$. The results for carbon-nanotubes have a significant temperature coefficient with  $C_\text{T}=0.70 \ (\beta_\text{T}=0.28)$. As seen in fig.~\ref{Figure 5}, the results for carbon-nanotubes go along with large error-bars. In contrast to the temperature, the size dependence showed insignificant results for copper oxide and carbon-nanotubes, both with $\beta_\text{S}=0.03$. The large error-bars for copper oxide as seen in fig.~\ref{Figure 6} correspond to the small range of measured nanoparticle diameters. Alumina and titania exhibit significant regression coefficients at the same level with $C_\text{S}=0.45 \ (\beta_\text{S}=0.36)$ for alumina and $C_\text{S}=0.37 \ (\beta_\text{S}=0.35)$ for titania.

\section*{DISCUSSION}

The main problem in the field of nanofluids seems to be the large scatter in the measurements. Nanofluids with the apparently same characterizations (concentration, temperature, size, material) exhibit different measurement results depending on the publication. The use of different measuring techniques is not the only reason \cite{Bobbo2021}. There is a lack of understanding on how to measure the thermal conductivity of dispersions; even in the measurement of pure fluids, partially large scattering occurs \cite{Chirico2013}. In the case of concentration, the use of volume-based units by dispersing colloidal nanoparticles is less common than the use of weight-based units\cite{Chirico2013, Bobbo2021}. The determination of the nanoparticle size and homogeneous distribution should be measured in all cases. A simple reference to the manufacturers information is not sufficient, because the particles act differently in dispersion than when stored dry. Comparing measurements of the sizes (e.g. TEM and DLS measurements) should not only lead to more significant results, they would also give a hint to the dispersion state and the agglomeration of the particles \cite{Bobbo2021}. Therefore it is important to examine and develop comparable measuring techniques. A careful characterization of the nanoparticles properties could lead to verifiable measurements. 

The results shown in table~\ref{tab:1} and fig.~\ref{Figure 4} suggest that the thermal conductivity of nanofluids is in agreement with Maxwell's effective medium theory. The concentration mainly determines the thermal conductivity, the temperature and nanoparticle size have lower significant influence. The temperature effect is only significant for carbon-nanotubes and marginal for copper oxide (see fig.~\ref{Figure 5}). As already stated by Prasher et. al., the influence of temperature on the thermal conductivity coincides with agglomeration in the dispersion \cite{Prasher2006}. As seen in fig.~\ref{Figure 4}, the concentration coefficient for copper oxides exceed Maxwell's prediction. This suggests a non-homogeneous dispersion state in the theory of Hashin-Shtrikman \cite{Hashin1962, Eapen2010} and agrees with the significant temperature coefficient; copper oxides  are known to build large agglomerates \cite{Jadhav2011}. Percolation networks likely occur without the use of surfactants or other surface treatments. Indeed, the data for copper oxide based on the statistical regression used no surfactants at all. In the case of carbon-nanotubes, the temperature effect is significant while the concentration coefficient fits to Maxwell's theory. In many cases surfactants or surface functionalizations are used to ensure the homogeneous dispersion of the carbon-nanotubes, but surfactants can also lead to a decrease in thermal conductivity \cite{Kim2018}. Fig.~\ref{Figure 5} therefore shows that while the data for alumina, titania and copper oxide are similar, the temperature coefficient and the corresponding error-bars for carbon-nanotubes significantly differ from the other materials. Significant size effects occur only for alumina and titania in our statistical results, while the results for carbon-nanotubes and copper oxides are insignificant (see fig.~\ref{Figure 6}). These insignificant size effects mainly result for the limited variation of nanoparticle sizes in the data sets. For carbon-nanotubes the variation of the size is limited due to the shape. It is also unclear whether the size of carbon-nanotubes is to be defined by the diameter or the length of the tube \cite{Eatemadi2014}. The size coefficient and the standardized regression coefficients for alumina and titania are in comparable ranges. This supports that the thermal conductivity of the nanofluid increases with decreasing particle sizes as already published \cite{Khanafer2011}. However, the uncertainties of the measurements and the difficulties in the stabilization of the well-dispersed state, are not taken into account in the statistical analysis.

\section*{CONCLUSION}
With our statistical analysis we show that the thermal conductivity of nanofluids can be well described by Maxwell's effective medium theory. Only copper oxides exceeds this prediction with \cphiCuO, possibly due to agglomeration effects. We found significant influence for the temperature in carbon nanotubes (\ctCNT) and for the particle size in alumina (\csAl) and titania (\csTi). The main limiting factor is in general the lack of precise measurements and characterizations. Furthermore, the formation of agglomerates and percolation networks increases the viscosity of the dispersion, which makes it unfavourable for application. With careful and comparable measurements, further statistical analyses would allow to asses the full potential of nanofluids \cite{Tielke2021}. 






{\footnotesize

\printbibliography
}

\endmytext 

\end{document}